\documentclass[a4paper,fleqn]{cas-dc}

\usepackage[authoryear]{natbib}

\def\tsc#1{\csdef{#1}{\textsc{\lowercase{#1}}\xspace}}
\tsc{WGM}
\tsc{QE}
\tsc{EP}
\tsc{PMS}
\tsc{BEC}
\tsc{DE}

\begin{document}
\let\WriteBookmarks\relax
\def\floatpagepagefraction{1}
\def\textpagefraction{.001}
\shorttitle{Detecting SARS-CoV-2 Remotely Using Recorded Cough Sounds.}
\shortauthors{L. H. Nguyen and N. T. Pham et~al.}

\title [mode = title]{Fruit-CoV: An Efficient Vision-based Framework for Speedy Detection and Diagnosis of SARS-CoV-2 Infections Through Recorded Cough Sounds (a Short Report)}                      
\tnotemark[1,2,3]

\tnotetext[1]{The proposed method in this study won the 1$^{st}$ place on the leaderboard of the AICovidVN 115M Challenge (https://www.covid.aihub.vn/).}

\tnotetext[2]{The study is conducted by the FruitLab team.}

\tnotetext[3]{A full version is in preparation and will be updated in an early date.}

\author[1]{Long H. Nguyen}[]
\cormark[1]
\ead{hoanglong.fruitai@gmail.com}
\address[1]{Faculty of Information Technology, Ton Duc Thang University, Ho Chi Minh City, Vietnam}
\credit{Propose ideas, develop software, and join the challenge}

\author[2]{Nhat Truong Pham}[orcid=0000-0002-8086-6722]
\cormark[1]
\ead{phamnhattruong.st@tdtu.edu.vn}
\address[2]{Division of Computational Mechatronics, Institute for Computational Science, Ton Duc Thang University, Ho Chi Minh City, Vietnam}
\credit{Propose ideas, draft the paper, and join the challenge}

\author[3]{Van Huong Do}[]
\ead{vanhuong.do@asicland.com}
\address[3]{ASICLAND, Suwon, South Korea}
\credit{Develop software and deploy the service}

\author[1]{Liu Tai Nguyen}[]
\ead{nguyenliutientai@gmail.com}
\address[1]{Faculty of Information Technology, Ton Duc Thang University, Ho Chi Minh City, Vietnam}
\credit{Develop software and deploy the service}

\author[4]{Thanh Tin Nguyen}[]
\ead{nttin@sju.ac.kr}
\address[4]{Human Computer Interaction Lab, Sejong University, Seoul, South Korea}
\credit{Revise the paper and support the development}

\author[5]{Van Dung Do}[]
\ead{dungdv1698@gmail.com}
\address[5]{Hanoi University of Industry, Hanoi, Vietnam}
\credit{Revise the paper and support the development}

\author[6]{Hai Nguyen}[]
\ead{hainguyen@ccs.neu.edu}
\address[6]{Khoury College of Computer Science, Northeastern University, Boston, USA}
\credit{Revise the paper and support the development}

\author[7]{Ngoc Duy Nguyen}
\cormark[2]
\ead{n.nguyen@deakin.edu.au}
\address[7]{Institute for Intelligent Systems Research and Innovation, Deakin University, Victoria, Australia}
\credit{Supervise the team, provide methodology and appropriate guidance, and revise the paper}

\cortext[cor1]{Co-first authors: these authors have contributed equally to this work.}
\cortext[cor2]{Corresponding author.}

\begin{abstract}
%
%
%
%
%
SARS-CoV-2 is colloquially known as COVID-19 that had an initial outbreak in December 2019. The deadly virus has spread across the world, taking part in the global pandemic disease since March 2020. In addition, a recent variant of SARS-CoV-2 named Delta is intractably contagious and responsible for more than four million deaths over the world. Therefore, it is vital to possess a self-testing service of SARS-CoV-2 at home. In this study, we introduce Fruit-CoV, a two-stage vision framework, which is capable of detecting SARS-CoV-2 infections through recorded cough sounds. Specifically, we convert sounds into Log-Mel Spectrograms and use the EfficientNet-V2 network to extract its visual features in the first stage. In the second stage, we use 14 convolutional layers extracted from the large-scale Pretrained Audio Neural Networks for audio pattern recognition (PANNs) and the Wavegram-Log-Mel-CNN to aggregate feature representations of the Log-Mel Spectrograms. Finally, we use the combined features to train a binary classifier. In this study, we use a dataset provided by the AICovidVN 115M Challenge, which includes a total of 7371 recorded cough sounds collected throughout Vietnam, India, and Switzerland. Experimental results show that our proposed model achieves an AUC score of 92.8\% and ranks the 1$^{st}$ place on the leaderboard of the AICovidVN Challenge. More importantly, our proposed framework can be integrated into a call center or a VoIP system to speed up detecting SARS-CoV-2 infections through online/recorded cough sounds.
%
%
\end{abstract}

%


\maketitle

\section{Introduction}

Delta is an instantly contagious variant of SARS-CoV-2 that has caused a multi-continent disaster. Specifically, a study from the World Health Organization (WHO) reports that the COVID-19 pandemic is responsible for 4,400,284 deaths and 209,876,613 confirmed cases over the world until August 2021~\footnote{https://covid19.who.int/}. In addition to the traditional COVID-19 testing methods that require a physical presence of patients, it is crucial to complement a free-of-charge self-testing service from home with an instant outcome. In this study, we investigate a method of detecting and diagnosing SARS-CoV-2 infections through cellphone recorded cough sounds.

In particular, we propose a vision-based method that transforms a sound into a Log-Mel Spectrogram. We then introduce a two-stage deep neural architecture and employ a supervised training on the network to classify between infected or non-infected people. The first stage includes the baseline EfficientNet-V2 (EffNetV2) network that extracts visual features of the Log-Mel Spectrograms. In the second stage, we use the Wavegram-Log-Mel CNN, followed by 14 convolutional layers extracted from the PANNs to further process feature representations of the Log-Mel Spectrograms. The aggregated features are then used to train the network to detect infected cough sounds.

In this study, we use a dataset provided by the AICovid-VN 115M Challenge to evaluate the efficiency of the proposed framework. The dataset combines two public datasets from India and Switzerland and a collection of recorded cough sounds in Vietnam. As a result, the dataset includes 4,508 training sounds, 1,230 public testing sounds, and 1,627 private testing sounds. Our proposed framework achieves an AUC score of 92.8\% in a private test.  

The rest of this paper is organized as follows. Related studies are summarized in section~\ref{sect2}. Fruit-CoV is presented in section~\ref{sect3}. In section~\ref{sect4}, we present the evaluation results. Finally, section~\ref{sect5} concludes the paper.

%
%
%

\section{Related Work} \label{sect2}
Researchers have a great interest in employing low-cost applications toward detecting SARS-CoV-2 through cough and breath sounds. For instance, the authors in \citep{pub1} used transfer learning to develop an AI-based diagnostic system using recorded cough sounds. In particular, the authors extracted Mel Frequency Cepstral Coefficients (MFCCs) from sounds to train a classifier based on three pre-trained ResNet50s and a Poisson biomarker layer. The model achieves an accuracy of 95\% and 99.9\% on groups of 25 people with five and three positive cases, respectively.  

The authors in~\citep{pub8} adopted a machine learning binary classifier to identify the presence of COVID-19 using cough and breath sounds in a crowdsourced respiratory dataset. In \citep{pub10}, the authors used a wide range of machine learning and deep learning techniques to detect infected cough sounds. For instance, logistic regression (LR), support vector machine (SVM), multilayer perceptrons (MLP), convolutional neural networks (CNN), long short-term memory (LSTM), and ResNet50 were used. SMOTE oversampling was adopted to the minor class, which accounts for 7.86\% of cough sounds. The authors in~\citep{pub12} used both unsupervised and supervised learning approaches to analyze different types of cough sounds, e.g., dry, wet, whooping, and COVID-19 coughs. However, the method only uses raw audio signals from a limited dataset. An ensemble learning technique as in~\citep{pub30} consists of a shallow MLP, a CNN, and a pre-trained CNN. However, the accuracy of this technique is not greater than 90\%.

Many datasets of cough sounds have been created to encourage the research community into the combat with SARS-CoV-2. For example, the authors in \citep{pub11} created a COVID-19 dataset, namely Coswara, which comprises breathing, coughing, and voice sounds. The dataset includes 6,507 clean recorded sounds and 1,117 noisy ones collected from 941 people. Additionally, the authors in \citep{pub4} created a dataset of 73 samples to complement a lack number of COVID-19 datasets. Another dataset named COUGHVID was provided to the research community by \citep{pub24}. This dataset contains approximately 20,000 recorded cough sounds accompanying the bioinformation such as age, gender, geographic location, and COVID-19 positive/negative testing outcome.

\section{Proposed Method} \label{sect3}

\subsection{Dataset}
The AICovidVN 115M Challenge is a well-known community project in Vietnam toward finding an efficient method of testing COVID-19 based on cough sounds. The winning team is selected to deploy the testing service on a broader scale in Vietnam. The dataset contains two classes, i.e., the positive label and the negative label. It comprises 4,508 training and 2,863 testing samples. A wide array of different devices are utilized to record the sounds.





\subsection{Fruit-CoV}
In this paper, we used the baseline EfficientNet-V2 in the first stage to extract high-level visual representations from the Log-Mel Spectrograms. In the second stage, a pre-trained CNN with 14 layers from the PANNs is used to extract additional embedding features of the Log-Mel Spectrograms. Because the PANNs was trained on a large-scale AudioSet dataset that consists of many respiratory audio classes, the extracted features can be used as bio-embedding features. Embedding features of two branches are concatenated to eventually train a binary classifier. Figure~1 describes an overview of our proposed scheme.

\begin{figure*}[h!]
	\centering
	\includegraphics[width=0.95\textwidth]{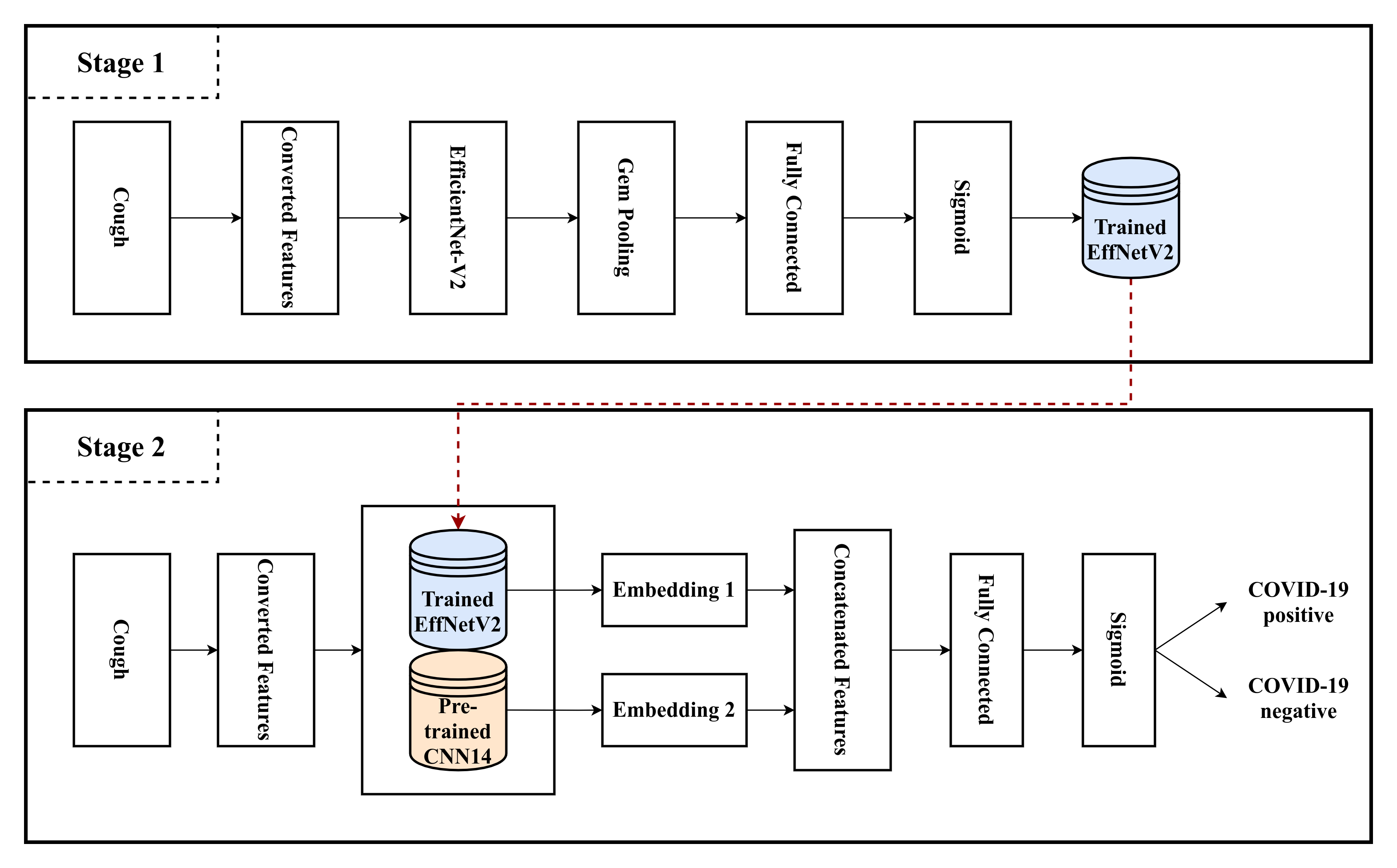}
	\caption{The proposed scheme of two-stage framework.} \label{fig1}
\end{figure*}

The dataset includes three major sampling rates: 4 kHz, 8 kHz, and 48 kHz. We apply the following data processing:

\begin{itemize}
	\item In the case of 4 kHz: we convert cough sounds into Log-Mel Spectrograms with a sampling rate of 4 kHz. The Log-Mel Spectrograms are used as inputs of both stages. In the first stage, besides EfficientNet-V2, a GeM Pooling is added to seek for important feature representations. Such representations are propagated into a fully connected layer to classify infected or non-infected cases. In the second stage, sounds are converted into Log-Mel Spectrograms using a sampling rate of 8 kHz before feeding into the CNN14. Moreover, to adapt with the embedding features (Embedding 1) extracted from the trained EffNetV2, we select the features from the `conv\_block6' of the CNN14, followed by a GeM Pooling layer to generate the embedding features, i.e., Embedding 2.
	\item In the case of 8 kHz: we apply a similar procedure as of 4 kHz except the Log-Mel Spectrograms are processed by a sampling rate of 8 kHz in both two stages. Additionally, the Embedding 2 is chosen from the `embedding' layer of the CNN14 without adding the GeM Pooling layer.
	\item In the case of 48 kHz: The procedure is different from the first two cases. Specifically, we convert cough sounds into Log-Mel Spectrograms with a sampling rate of 48 kHz in the first stage. However, in the second stage, to avoid losing the quality of sounds and unsynchronizing with the pre-trained model's weights, we feed the Wavegram-Log-Mel-CNN sounds with a sampling rate of 32 kHz before applying the CNN14 to generate the Embedding 2.
	\item For other sampling rates, we convert sounds into Log-Mel Spectrograms using the closest sampling rate as listed in case 1, case 2, or case 3.
\end{itemize}


\section{Performance Evaluation} \label{sect4}

\subsection{Hyperparameter Settings}

Hyperparameters are shown in Table~1, where the SpecAugment augmentation method is applied to train the baseline, i.e. EffNetV2, in the first stage of our proposed framework. Additionally, we use the k-fold cross-validation method to generalize the model with $k = 5$.

\begin{table}[h!]
\centering
\label{tab1}
\caption{The hyper-parameter settings of our proposed framework.}
\begin{tabular}{l|c}
\hline
\textbf{Params}         & \textbf{Configure/Setup} \\ \hline
Learning rate           & $e^{-3}$                   \\ \hline
Learning rate scheduler & Cosine ($e^{-5}$)          \\ \hline
Batch size              & 16                       \\ \hline
Epochs                  & 30                       \\ \hline
Optimization            & CosAngularGrad           \\ \hline
Loss                    & BCE with logits          \\ \hline
Augmentation            & SpecAugment              \\ \hline
\end{tabular}
\end{table}


To convert cough sounds into Log-Mel Spectrograms, we apply a Mel bin of 256 in the case of 4 kHz, while other cases use a Mel bin of 128. Librosa is used to convert cough sounds into Log-Mel Spectrograms in this study.

\subsection{Evaluation Results}

Figure~2 shows AUC scores of the first stage with different sampling rates while AUC scores of the second stage are shown in Figure~3. In Figure~2, the scores reach 93\%, 96\%, and 94\% in the first stage with sampling rates of 4 kHz, 8 kHz, and 48 kHz, respectively. Similarly, in the second stage, the scores reach 95\%, 97\%, and 99\%.

\begin{figure*}
    \centering
	\includegraphics[width=0.32\textwidth]{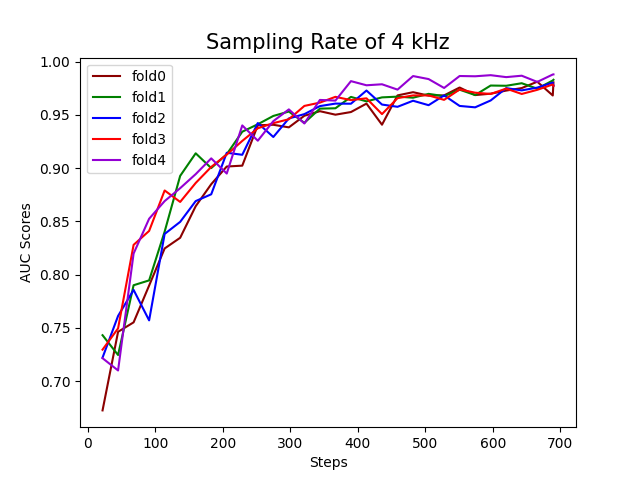}
	\includegraphics[width=0.32\textwidth]{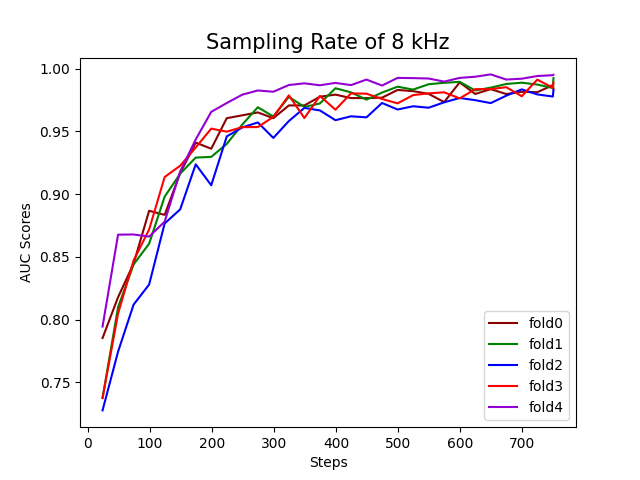}
	\includegraphics[width=0.32\textwidth]{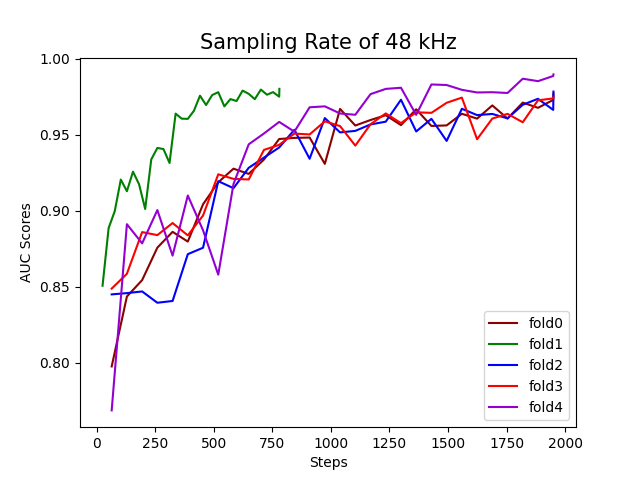}
	
	\caption{AUC Scores on the first stage with 4, 8, and 48 kHz, respectively (from left to right).}
	\label{fig:gem}
\end{figure*}

\begin{figure*}
	\centering
	\includegraphics[width=0.32\textwidth]{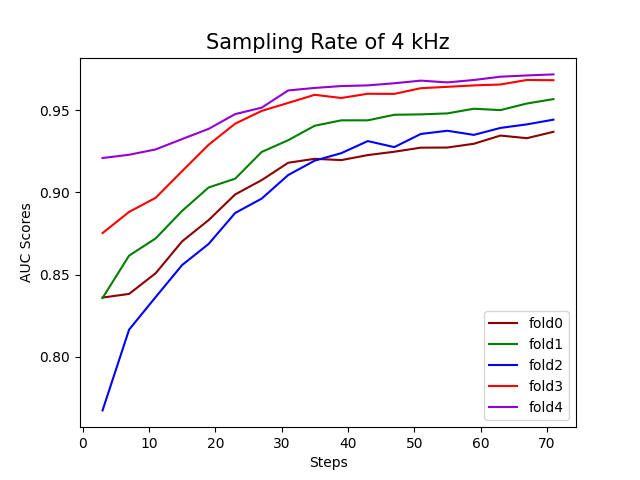}
	\includegraphics[width=0.32\textwidth]{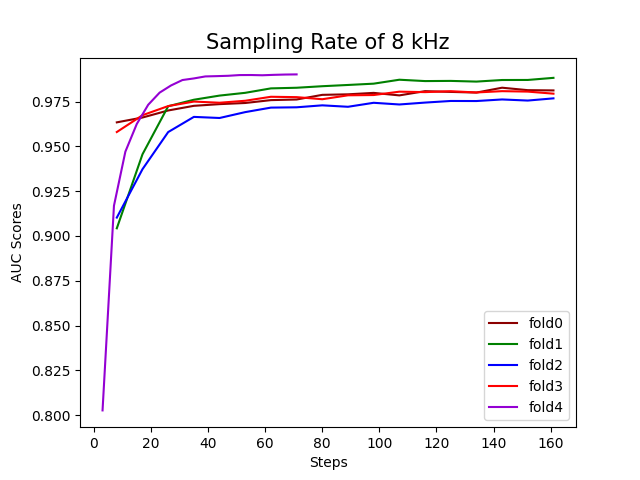}
	\includegraphics[width=0.32\textwidth]{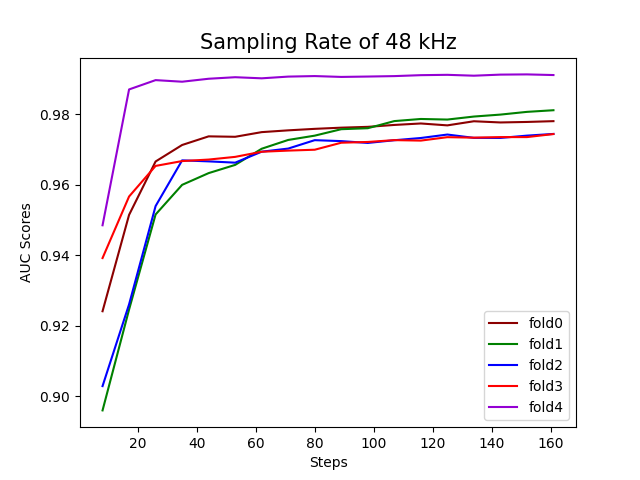}
	\label{fig:hybrid}
	\caption{AUC Scores on the second stage with 4, 8, and 48 kHz, respectively (from left to right).}
\end{figure*}

Comparing to existing studies, our proposed method is more flexible with different sampling rates. The AUC score reaches 92.8\% on the private test of the AICovidVN 115M Challenge. 


\section{Conclusion} \label{sect5}
This study proposed a vision-based framework, namely Fruit-CoV, which provides high performance in detecting and diagnosing SARS-CoV-2 infections through recorded cough sounds. We introduced a hybrid deep neural network architecture that utilizes state-of-the-art architectures, i.e., EfficientNet-V2 and PANNs. We also meticulously analyze the dataset and apply appropriate data preprocessing. Experimental results show that our proposed Fruit-CoV framework attains an AUC score of 92.8\% and won the 1$^{st}$ place on the leaderboard of the AICovidVN 115M Challenge. In addition, Fruit-CoV can be integrated into a call center, a VoIP system, or a mobile application to deploy a self-testing service of COVID-19 from home rapidly. We continuously improve the Fruit-CoV framework by aggregating recorded cough sounds, automating the data processing, exploring other state-of-the-art techniques, and generating an evaluation benchmark of existing methods.



\section*{Acknowledgment}

The authors would like to thank the organizer of the AI-Covid-VN 115M Challenge for providing the recorded cough sound dataset.

\printcredits

\bibliographystyle{cas-model2-names}



%
%

\end{document}